\documentclass{iaus}
\usepackage{graphics}

  \checkfont{eurm10}
  \iffontfound
    \IfFileExists{upmath.sty}
      {\typeout{^^JFound AMS Euler Roman fonts on the system,
                   using the 'upmath' package.^^J}%
       \usepackage{upmath}}
      {\typeout{^^JFound AMS Euler Roman fonts on the system, but you
                   dont seem to have the}%
       \typeout{'upmath' package installed. iaus.cls can take advantage
                 of these fonts,^^Jif you use 'upmath' package.^^J}%
      }
  \else
  \fi

  \checkfont{msam10}
  \iffontfound
    \IfFileExists{amssymb.sty}
      {\typeout{^^JFound AMS Symbol fonts on the system, using the
                'amssymb' package.^^J}%
       \usepackage{amssymb}%

      }{}
  \fi

  \IfFileExists{amsbsy.sty}
    {\typeout{^^JFound the 'amsbsy' package on the system, using it.^^J}%
     \usepackage{amsbsy}}
    {}

\newsavebox{\astrutbox}
\sbox{\astrutbox}{\rule[-5pt]{0pt}{20pt}}

\title[The Interplay among Black Holes, Stars and ISM in Galactic 
       Nuclei]{The \vcirc -\sigc\ relation in high and low surface brightness
galaxies}

\author[A. Pizzella {\it et al.\/}]%
{Alessandro Pizzella$^1$%
E. Dalla Bont\`a$^1$
E. M. Corsini$^1$
L. Coccato$^1$
F. Bertola$^1$
\and M. Sarzi$^2$}

\affiliation{$^1$Dipartimento di Astronomia, Universit\`a di Padova,
Padova, Italy\\[\affilskip]
$^2$University of Oxford, UK}

\pubyear{2004}
\volume{222}
\pagerange{1--8}
\date{?? and in revised form ??}
\jname{The Interplay among Black Holes, Stars and ISM \\in Galactic Nuclei}
\editors{Th. Storchi Bergmann, L.C. Ho \& H.R. Schmitt, eds.}
\def\vcirc{$V_{\rm circ}$}
\def\sigc{$\sigma_{\rm c}$}

\begin{document}

\maketitle

\begin{abstract}
We investigate the relation between the asymptotic circular velocity,
\vcirc , and the central stellar velocity dispersion, \sigc , in
galaxies.  We consider a new sample of high surface brightness spiral
galaxies (HSB), low surface brightness spiral galaxies (LSB), and
elliptical galaxies with HI-based \vcirc\ measurements.  
We find that:\\ 
1) elliptical galaxies with HI measurements fit well within the
relation;\\
2) a linear law can reproduce the data as well as a power law (used in
 previous works) even for  galaxies with \sigc$< 70$ km/s;\\
3) LSB galaxies, considered for the first time with this respect, seem
to behave differently, showing either larger \vcirc\ values or
smaller \sigc\ values.
\end{abstract}

Recently a tight correlation between the bulge velocity dispersion
\sigc\ and the galaxy asymptotic circular velocity \vcirc\ has been
found for a sample of elliptical and spiral galaxies 
(\cite[Ferrarese 2002]{Ferr2002}). 
The validity of this relation has been also confirmed by \cite{Baes2003}, 
who enlarged the spiral galaxy sample.  The fact that
such a tight relation exists between two velocity scales that probe
very different spatial regions (the bulge and the dark matter halo),
is a strong indication of a fundamental correlation in the structure
not only of spirals but also of ellipticals.  On the other hand, it
may be interesting to investigate whether the \vcirc -\sigc\ relation
holds also for less dense objects characterized by a shallow potential
well in their core. This is the case of LSB galaxies.

We studied the \vcirc -\sigc\ adding to previous studies data for
HSBs, Es, and LSBs. In particular, we consider a sample of 41 HSB
spirals (17 from \cite{Ferr2002}, 7 from \cite[Baes et al. 2003]{Baes2003}, 17 from
\cite{Pizz2004}, 11 LSB spirals \cite{Pizz2003}, 19 Es
from \cite{Kron2000}, which \vcirc\ are based on dynamical
models. To check whether elliptical galaxies with model-independent
\vcirc\ values would still follow the same \vcirc -\sigc\ relation as
Spirals, we added 5 Es with HI measurements. 

We check that all galaxies have extended rotation curves which are
flat in the last region, in order to derive the value of \vcirc. The
resulting \vcirc -\sigc\ diagram is plotted in Fig. 1a  and the
conclusions of our work are summarized in the abstract. Here we point
out that we have at the moment data for only 11 LSB galaxies.
Although a KS test applied to the two distributions shown in Fig.1b,
indicates that they are different at a $3\sigma$ confidence level
 and thus that LSBs do not follow the same \vcirc -\sigc\
relation as HSB and E galaxies, we need more data points for LSB
galaxies to confirm such discrepancy.

Confirming this result will highlight yet another aspect in the
different formation history of LSBs. Indeed, LSBs appear to have a
central potential well less deep than HSB spirals of the same halo
mass. If the collapse of baryonic matter cause a compression of the
dark halo as well, for LSB galaxies such process may have been less
relevant than for HSBs.  Again LSBs turn out to be the best tracers of
the original density profile of dark matter halos and therefore in
pursuing the nature of dark matter itself.


\begin{figure}
 \scalebox{0.35}{\includegraphics{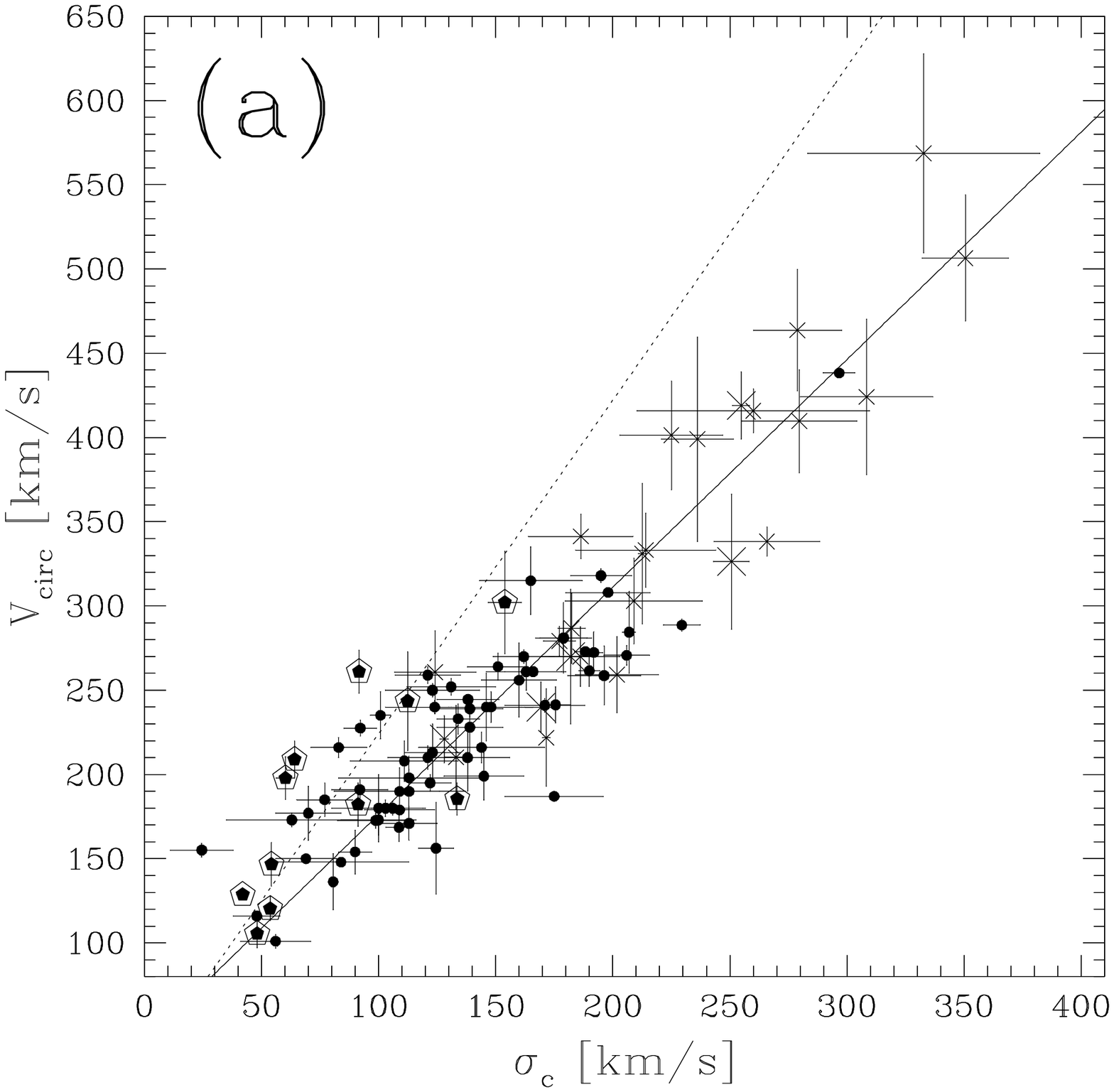}}
 \scalebox{0.35}{\includegraphics{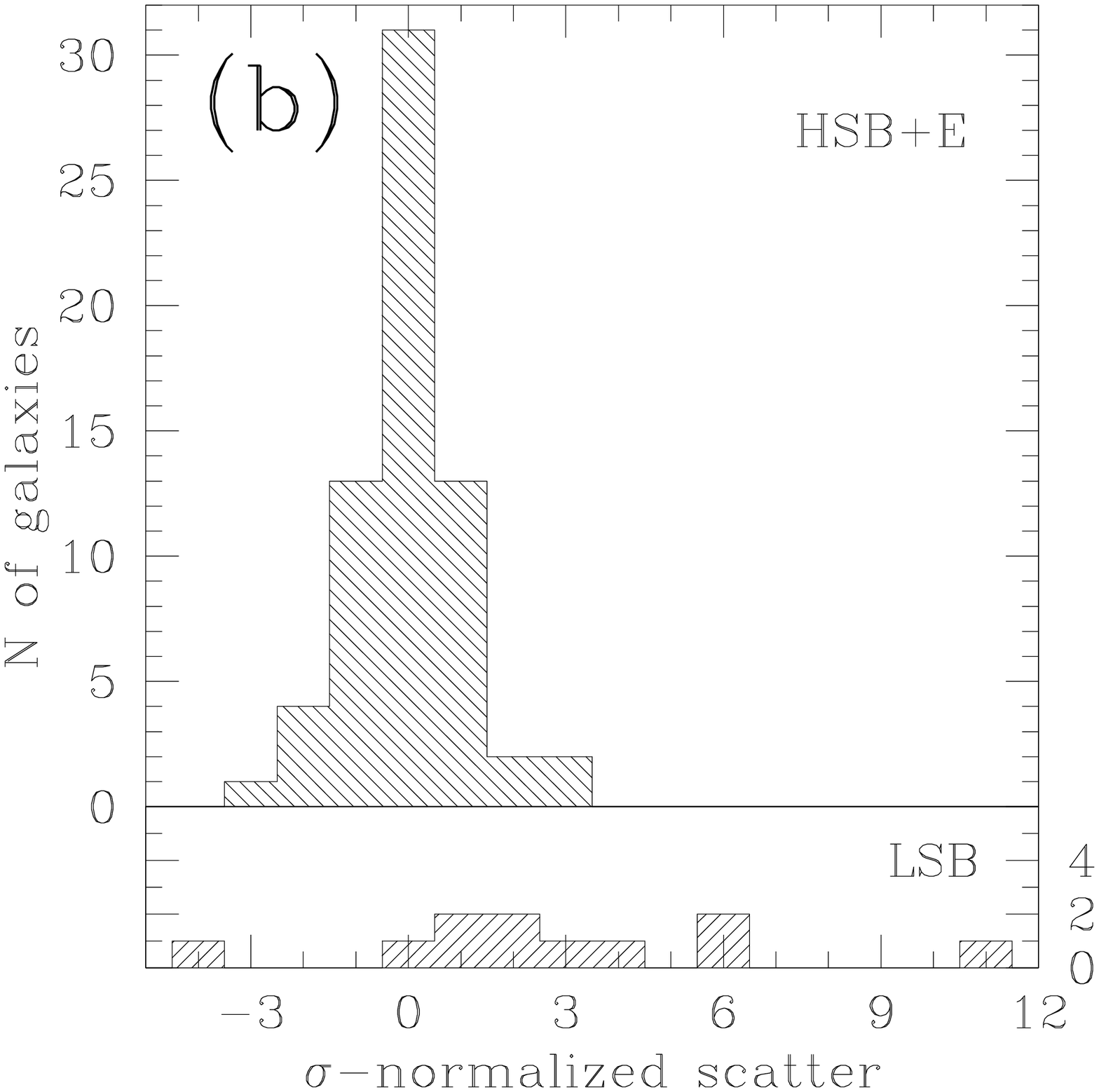}}

\caption{{\bf (a)}: \vcirc -\sigc\ relation for galaxies, including
ellipticals from Kronawitter et al. ({\it crosses}) or with HI
measurements ({\it big crosses}), HSBs ({\it dots}), and LSBs ({\it
dotted pentagons}). The {\it Full line} represents the linear
regression fit of the HSB+E sample. The {\it dotted line} is the fit
on the LSB sample.
{\bf (b)}: distribution of the scatter of the data-points with respect
to the HSB+E \vcirc -\sigc\ relation ({\it Full line} of
Fig. 1a). The scatter accounts for the error bar of each data
point. In the {\it upper panel} we plot the histogram relative to the
HSB+E sample and in the {\it lower panel} the histogram relative to
the LSB sample. A KS test indicates that the two
distributions are different at a confidence level higher than $3\sigma$.}
\end{figure}

%
%

\end{document}